\documentclass[twoside,11pt]{article}

\usepackage{jmlr2e}
\usepackage{booktabs}
\usepackage{lastpage}
\usepackage{xcolor}
\definecolor{okgreen}{HTML}{1B7A3D}
\newcommand{\yes}{\textcolor{okgreen}{\textbf{yes}}}

\makeatletter
\long\def\@makecaption#1#2{%
   \vskip 3pt
   {\scriptsize
   \setbox\@tempboxa\hbox{#1: #2}%
   \ifdim \wd\@tempboxa >\hsize
       \begin{list}{#1:}{%
       \settowidth{\labelwidth}{#1:}%
       \setlength{\leftmargin}{\labelwidth}%
       \addtolength{\leftmargin}{\labelsep}%
        }\item #2 \end{list}\par
     \else
       \hbox to\hsize{\hfil\box\@tempboxa\hfil}%
   \fi}}
\makeatother

\makeatletter
\def\section{\@startsiction{section}{1}{\z@}{-0.10in}{0.03in}
             {\large\bf\raggedright}}
\makeatother
\long\def\acks#1{\vskip 0.04in\noindent{\large\bf Acknowledgments and
Disclosure of Funding}\vskip 0.02in\noindent #1}

\jmlrheading{27}{2026}{1-\pageref{LastPage}}{}{}{}{Sankalp Gilda}

\ShortHeadings{tsbootstrap: Distribution-Free UQ for Time Series}{Gilda}
\firstpageno{1}

\begin{document}

\title{tsbootstrap: Distribution-Free Uncertainty Quantification and
Conformal Prediction for Time Series}

\author{\name Sankalp Gilda \\
       \addr DeepThought Solutions \\ ORCID: 0000-0002-3645-4501}

\editor{}

\maketitle

\begin{abstract}%
Finance, sensing, and demand streams violate the exchangeability that IID
conformal prediction and the IID bootstrap assume, and existing libraries
implement either a general resampling engine or conformal calibration
without the other.
\texttt{tsbootstrap} provides block, residual, sieve, and wild resampling,
classical bootstrap confidence intervals, and adaptive conformal calibrators
(EnbPI, ACI, NexCP, AgACI) through a single typed API in which a
specification object selects each method. In a controlled coverage study the
IID bootstrap undercovers sharply under dependence; dependence-aware methods
reduce the coverage deficit, the sieve nearest to nominal under
short-memory linear dependence. On the shared fixed-statistic path a compiled backend
runs several times faster than \texttt{arch}, and a streaming reduce avoids
materializing the $O(Bn)$ replicate tensor, limiting peak extra memory to $O(B)$
for the statistic array. The software is MIT licensed (v0.6.1).
\end{abstract}

\begin{keywords}
  conformal prediction, uncertainty quantification, bootstrap, time series,
  distribution-free inference
\end{keywords}

\section{Introduction and Statement of Need}

Practitioners pair point forecasts with calibrated intervals, but the two
distribution-free tools that would supply one assume away temporal structure:
split conformal prediction assumes exchangeable calibration data, the
ordinary bootstrap assumes IID observations, and both undercover on
autoregressive streams.
The time series bootstrap restores validity under dependence, and conformal
calibration supplies the finite-sample coverage guarantee.
Prior libraries supply one piece: forecasting toolkits (\texttt{skforecast},
\texttt{darts}, \texttt{statsforecast}, \texttt{neuralforecast}) attach
residual-bootstrap or split-conformal intervals to their own forecasters,
conformal libraries (\texttt{MAPIE}, \texttt{TorchCP}, \texttt{puncc})
calibrate a fitted predictor, and resampling libraries (\texttt{arch}) stop
at classical intervals. To our knowledge, no existing library combines a
general dependence-aware resampling engine (block, residual, sieve, and wild
methods with typed specs, for arbitrary statistics) with an adaptive
conformal layer and a streaming reduce in one API;
\texttt{tsbootstrap} provides this combination. Derivations appear in
the companion methods manuscript (arXiv:2404.15227).

\section{Design and API}

The single typed entry point \texttt{bootstrap(X, *, method=spec, ...)}
selects the method by a concrete specification object whose \emph{type}
identifies it, visible to editors and type checkers. A typical analysis
consists of three calls: \texttt{diagnose(X)} to recommend a spec,
\texttt{bootstrap(X, method=spec, n\_bootstraps=999, random\_state=0)}, and
\texttt{conf\_int(res, statistic="mean")}.

The call returns a structured \texttt{BootstrapResult}:
\emph{samples} (the replicate array), \emph{provenance} (spec, seed,
backend), \emph{out-of-bag} (the held-out index EnbPI calibrates on), and
\emph{in-bag} (the drawn index, for jackknife-after-bootstrap and studentized
errors). Inputs are handled by a \texttt{narwhals} layer (a lightweight
compatibility layer over dataframe libraries): NumPy arrays, Python lists,
and frames or series from \texttt{pandas}, \texttt{Polars}, and
\texttt{PyArrow}.
Each spec carries a \texttt{MethodMetadata} record
(\texttt{metadata\_for(spec)}) declaring assumptions, capabilities,
references, cost, and failure modes; \texttt{diagnose(X)} reads the lag-one
autocorrelation and an ADF stationarity test and returns a frozen
\texttt{Diagnosis} naming recommended specs with an auto-selected block
length.

Results are deterministic per backend for a fixed seed: each replicate draws
from its own RNG stream, so output is invariant to replicate ordering,
parallelism, and thread count. The default \texttt{numpy} backend spawns
per-replicate PCG64 streams via \texttt{SeedSequence} and is bit-for-bit
reproducible for a pinned seed and environment. The optional compiled \texttt{numba} backend
(\texttt{[accel]} extra), the source of the Section~\ref{sec:perf}
throughput, draws from a distinct counter-based stream, equal in distribution
but not bit-identical to the \texttt{numpy} path; its byte stream is pinned
within a release, not across releases.

\section{Methods and UQ Coverage}

Block methods cover moving \citep{kunsch1989}, circular \citep{politis1992},
stationary \citep{politis1994}, non-overlapping \citep{carlstein1986}, and
tapered blocks with energy-normalized windows \citep{paparoditis2001}, with
block length defaulting to the automatic spectral-density plug-in rule of
\citet{politis2004}, with the correction of \citet{patton2009}.
Model-based bootstraps (\texttt{ResidualBootstrap} over AR, ARIMA, and VAR)
regenerate series \emph{recursively} from the fitted dynamics and resampled
centered innovations \citep{pascual2004}, rather than the less accurate
fitted-plus-residuals scheme; \texttt{SieveAR} implements the sieve
bootstrap \citep{buhlmann1997,kreiss1992}, with exogenous support (ARX, VARX,
ARIMAX). Innovations may be resampled IID
(default), wild (Rademacher by default)
\citep{wu1986,liu1988,mammen1993,davidson2008}, or block-wild for
dependent innovations \citep{shao2010}.

Classical intervals via \texttt{conf\_int}
cover percentile \citep{efron1979}, basic, studentized bootstrap-t with a
dependence-aware block-jackknife standard error \citep{kunsch1989}, and BCa
\citep{efron1987}, the last gated to the IID spec because the studentized
route is the second-order-correct choice under dependence \citep{gotze1996}.
The \texttt{uq} API adds conformal and adaptive intervals: EnbPI
out-of-bag intervals for sklearn-style regressors \citep{xu2021}, AR forecast
bands, and static, sliding-window, ACI \citep{gibbs2021}, NexCP
\citep{barber2023}, and AgACI \citep{zaffran2022} calibrators for stationary,
volatility-clustered, and drifting regimes. Table~\ref{tab:cov} reports
the empirical coverage these methods achieve.

\textbf{Coverage under dependence.}
The study crosses five DGPs, 5{,}000 Monte Carlo datasets each, eight
methods, and four functionals: 800{,}000 percentile intervals from $B{=}999$
replicates at $\alpha{=}0.10$; the table shows the mean for four DGPs and
four methods. DGPs: white noise $N(0,1)$; AR(1), $\phi{=}0.9$,
Gaussian innovations; an MA(2) control, $\theta{=}(0.6, 0.3)$ (not
displayed); AR(1)+ARCH(1), $x_t = 0.5x_{t-1} + \sigma_t\varepsilon_t$ with
$\sigma_t^2 = 0.2 + 0.4x_{t-1}^2$; ARFIMA($0, 0.4, 0$); series lengths in
the caption. The ARFIMA series are generated exactly by the Davies-Harte
circulant-embedding method; the white-noise series are drawn directly, the
MA(2) series by a direct causal filter, and the AR(1) and AR+ARCH series
recursively from their defining equations, with burn-ins of 400 (AR(1)),
200 (MA(2)), and 300 (AR+ARCH).
Coverage targets are exact-analytic from each DGP's closed-form
autocovariance (for AR+ARCH via its martingale-difference property),
corroborated by independent Monte Carlo to within 1\%. Block lengths use the
automatic rule above; the sieve selects its order by BIC; the
design is paired (one shared bootstrap seed per dataset across methods); the
maximum Monte Carlo standard error is 0.7 percentage points.
Dependence-aware methods correct much of the IID bootstrap's undercoverage:
on AR(1) with $\phi{=}0.9$ at $n{=}200$, coverage recovers from 27.8\% to
roughly 70\% (blocks) and 83.1\% (sieve), nearest to nominal under short-memory linear
dependence; the blocks fall short because the dependence is strong relative
to the series length. Under AR(1)+ARCH(1) the sieve reaches 89.1\%, within a
point of nominal. No method considered attains acceptable coverage under
long-memory ARFIMA ($d{=}0.4$), so the library exposes the method choice
rather than imposing a single default.

\begin{table}[ht]
\centering\scriptsize
\renewcommand{\arraystretch}{0.9}
\begin{tabular}{@{}lcccc@{}}
\toprule
DGP (nominal 90\%) & IID & Moving & Stationary & Sieve \\
\midrule
White noise      & 89.6 & 89.0 & 88.6 & 88.6 \\
AR(1)            & 27.8 & 69.7 & 70.2 & \textbf{83.1} \\
AR + ARCH        & 65.1 & 85.2 & 84.8 & \textbf{89.1} \\
ARFIMA (long memory) & 8.2 & 27.1 & 28.9 & 21.9 \\
\bottomrule
\end{tabular}
\caption{Empirical coverage (\%) of the bootstrap mean interval at 90\%
nominal ($n{=}200$ white noise and AR(1), $n{=}500$ AR+ARCH, $n{=}1000$
ARFIMA; 5{,}000 Monte Carlo datasets per cell; max Monte Carlo standard
error 0.7pp). The sieve is nearest to nominal under short-memory linear
dependence; no method reaches acceptable coverage under ARFIMA.}
\label{tab:cov}
\end{table}

\section{Comparison}

Table~\ref{tab:compare} compares the five nearest libraries;
\texttt{tsbootstrap} is the only column with entries in both the resampling
and conformal rows. \texttt{TorchCP} \citep{huang2024torchcp} and
\texttt{MAPIE} target post-hoc conformal calibration, \texttt{tsbootstrap}
dependence-aware uncertainty quantification. Because the adaptive calibrators
are pure functions over user-supplied nonconformity scores, residuals from any
prediction pipeline can be calibrated. Beyond
the table, \texttt{puncc} ships EnbPI; \texttt{statsforecast} and
\texttt{neuralforecast} attach split-conformal intervals to their own
forecasters; none has a general resampling engine.

\begin{table}[ht]
\centering\scriptsize
\setlength{\tabcolsep}{3.5pt}
\renewcommand{\arraystretch}{0.9}
\begin{tabular}{@{}lcccccc@{}}
\toprule
Capability & \textbf{tsbootstrap} & \texttt{arch} & \texttt{MAPIE} & \texttt{TorchCP} & \texttt{darts} & \texttt{skforecast} \\
\midrule
Dependence-aware TS resampling & \yes & \yes & partial$^{\dagger}$ & no & no & partial$^{\ddagger}$ \\
Wild / block-wild innovations & \yes & no & no & no & no & no \\
Classical bootstrap CIs & \yes & \yes & no & no & no & no \\
Adaptive conformal calibrators & \yes & no & partial & partial & no & no \\
Recommender + method metadata & \yes & no & no & no & no & no \\
Streaming bounded-memory reduce & \yes & \yes & no & no & no & no \\
Multivariate VAR + ragged panel & \yes & partial & no & no & partial & no \\
\texttt{sktime} / \texttt{skbase} adapters & \yes & no & no & no & no & no \\
\bottomrule
\end{tabular}
\caption{Capability comparison; conservative entries where support is
limited rather than absent. Versions (source-verified 2026-07-06):
\texttt{arch} 8.0, \texttt{MAPIE} 1.4.1, \texttt{TorchCP} 1.2.1,
\texttt{darts} 0.45.0, \texttt{skforecast} 0.23.0, \texttt{puncc} 0.9.3,
\texttt{statsforecast} 2.0.3, \texttt{neuralforecast} 3.1.9;
\texttt{tsbootstrap} 0.6.1. $^{\dagger}$\texttt{MAPIE}'s block bootstrap is
only an EnbPI calibration cross-validator.
$^{\ddagger}$\texttt{skforecast} resamples one-step residuals to simulate
forecast paths for its own forecasters (no block or sieve engine, no
arbitrary statistics); its conformal mode is split, without adaptive
calibrators. Adaptive row: \texttt{tsbootstrap} ships EnbPI, ACI, NexCP,
AgACI; \texttt{TorchCP} ships EnbPI, ACI, AgACI and \texttt{MAPIE} EnbPI,
ACI (for fitted predictors, not coupled to a resampling engine);
\texttt{darts} ships split and quantile wrappers only. NexCP is
\texttt{tsbootstrap}-only among the compared libraries.}
\label{tab:compare}
\end{table}

\section{Performance}
\label{sec:perf}

Our harness (\texttt{bench\_vs\_arch.py}) runs the compiled reducer and
\texttt{arch}'s \texttt{bs.apply(np.mean, reps=B)}, each library's own
streaming reducer and the closest like-for-like path, on the same warmed
workload on a dedicated Hetzner ccx33 (eight vCPU, AMD EPYC-Milan; Ubuntu
24.04, Python 3.12.3, NumPy 2.4.6, \texttt{arch} 8.0.0, \texttt{numba}
0.65.1), measured at tsbootstrap 0.4.0 (commit \texttt{53aa089}) and
confirmed directionally by a pinned v0.6.1 re-run. The comparison is not
symmetric in generality: \texttt{arch} pays a per-replicate Python callback
for an arbitrary statistic, while the compiled kernel fixes the statistic,
fuses the loop, and runs replicate-parallel on all eight threads against
\texttt{arch}'s single-threaded apply loop; the speedups below quantify this
fixed-statistic path only.

The sixteen-cell grid crosses the four methods both libraries implement
(IID, MovingBlock, CircularBlock, StationaryBlock) with
$n \in \{200, 2000\}$ and $B \in \{999, 10000\}$ on an AR(1) input
($\phi{=}0.6$), block length 20, mean statistic; each cell is the median of
three warmed repetitions, JIT and import excluded (full grid in
\texttt{benchmarks/}). The compiled backend is faster in all sixteen cells:
at $n{=}2000$ it runs $3.9\times$ faster on IID at $B{=}999$ (3.0 versus
11.7 ms) and $34.0\times$ faster on CircularBlock at $B{=}10000$ (4.9 versus
167.0 ms); the $n{=}200$ cells are larger still, from \texttt{arch}'s
per-replicate Python overhead; pinned to one thread it still leads by
roughly $2\times$ to $8\times$ (v0.6.1 re-measurement). For a
generality-matched comparison, the default \texttt{numpy} backend (which,
like \texttt{arch}, accepts an arbitrary statistic) runs $1.3\times$ to
$2.6\times$ slower than \texttt{arch} on the same cells; the compiled
speedup comes entirely from fusing a fixed statistic into the kernel.
On memory, \texttt{bootstrap\_reduce} avoids materializing the $O(Bn)$
replicate tensor (the \texttt{numpy} backend reduces in 2{,}048-replicate
chunks; the compiled kernel holds only the $(B, |\theta|)$ statistic array).
At $n{=}2000$, $B{=}50000$ (MovingBlock mean), peak
resident memory above the interpreter floor is 1,944 MB materializing every
replicate versus 20 MB streaming, about $96\times$ ($103\times$ at
$B{=}10000$; fresh subprocesses, warmed JIT,
\texttt{benchmarks/results/}). The streaming footprint grows only with $B$,
about 0.4 KB per replicate, not with $n \times B$.

\section{Reproducibility, Quality, and Availability}

\texttt{tsbootstrap} is MIT licensed (\texttt{github.com/astrogilda/tsbootstrap},
PyPI v0.6.1, Zenodo DOI 10.5281/zenodo.8226495) and supports Python 3.10--3.13 with
five core dependencies (\texttt{numpy}, \texttt{scipy}, \texttt{pydantic},
\texttt{scikit-base}, \texttt{narwhals}); \texttt{statsmodels}, \texttt{scikit-learn},
and \texttt{numba} are optional extras. Test coverage is 93\% overall with a
blocking 80\% per-file CI gate; GitHub Actions CI is green on
\texttt{main} across a 12-cell matrix (Python 3.10--3.13 on Linux, macOS, and Windows)
and further gates lint, zero-error \texttt{mypy} and \texttt{pyright}, executed
notebooks, and the docs build; a nightly mutation ratchet (about 1{,}100 mutants
over the engine and model core) fails on any new survivor. The core also ships as
\texttt{sktime}/\texttt{skbase} estimator adapters (validated by
\texttt{check\_estimator}) plus a read-only MCP server (\texttt{tsbootstrap-mcp},
stdio, \texttt{uvx}-runnable) for LLM agents; 14 CI-executed notebooks are at
\texttt{tsbootstrap.readthedocs.io}.

\acks{We thank co-maintainers Franz Kir\'aly and Benedikt Heidrich and the wider
\texttt{sktime} and \texttt{scikit-base} communities. AI assistance (Anthropic's
Claude, via the Claude Code CLI) supported code refactoring, test scaffolding, and
manuscript editing; the author made all design decisions and verified every
contribution against the tests and cited literature.}

\bibliographystyle{plainnat}

\end{document}